\documentclass [a4paper,12pt]{article}
\usepackage{amssymb}
\usepackage[pdftex]{graphicx}
\pdfoutput=1
\usepackage{amsmath}
\usepackage{fancybox}
\usepackage{color}
\usepackage{geometry}

\newcommand{\Slash}[1]{{\ooalign{\hfil/ \hfil\crcr$#1$}}\hspace{-0.5mm}} 
\newcommand{\be}{\begin{equation}}
\newcommand{\ee}{\end{equation}}
\newcommand{\bea}{\begin{eqnarray}}
\newcommand{\eea}{\end{eqnarray}} 
\begin{document}
\setlength{\baselineskip}{18pt}
\begin{titlepage}

\begin{flushright}
OCU-PHYS-543         
\end{flushright} 
\begin{flushright}
NITEP-111         
\end{flushright} 
\vspace{1.0cm}
\begin{center}
{\LARGE\bf The strong CP problem and higher dimensional gauge theories} 
\end{center}
\vspace{25mm}

\begin{center}
{\large
Yuki Adachi, C.S. Lim$^a$   
and Nobuhito Maru$^{b, c}$
}
\end{center}
\vspace{1cm}
\centerline{{\it
Department of Sciences, Matsue College of Technology,
Matsue 690-8518, Japan.}}

\centerline{{\it
$^{a}$
Tokyo Woman's Christian University, Tokyo 167-8585, Japan }} 

\centerline{{\it 
$^b$
Department of Mathematics and Physics, Osaka City University,
Osaka 558-8585, Japan}}

\centerline{{\it 
$^c$
Namubu Yoichiro Institute of Theoretical and Experimental Physics (NITEP), }} 
\centerline{{\it Osaka City University,  
Osaka 558-8585, Japan}}
%
%

\vspace{2cm}
\centerline{\large\bf Abstract}
\vspace{0.5cm}

We discuss a natural scenario to solve the strong CP problem in the framework of the higher dimensional gauge theory.  
An axion-like field $A_y$ has been built-in as the extra-space component of the higher dimensional gauge field. 
The coupling of $A_y$ with gluons is attributed to the radiatively induced ``Chern-Simons" (CS) term. We adopt a toy model with some unknown gauge symmetry U(1)$_{\rm X}$. The CS term is obtained in two ways: first by a concrete 1-loop calculation and next by use of the Fujikawa's method to deal with the chiral anomaly in 4D space-time. The obtained results are identical, which implies that the radiative correction to the CS term is ``1-loop exact" and is also free from UV-divergence even though the theory itself is non-renormalizable. As a novel feature of this scenario, such obtained CS term is no longer linear in the field $A_y$ as in the usually discussed CS term in 5D space-time but a periodic function of $A_y$, since $A_y$ has a physical meaning as the Wilson-loop phase. We argue how such novel feature of this scenario causes the modification of the ordinary solutions of the strong CP problem based on the axion fields.

\end{titlepage}

\section{Introduction} 

One of the attractive candidate of the physics beyond the standard model (BSM) is the scenario based on higher dimensional theories. 
Among them, higher dimensional gauge theory is of special interest. In the theory where the extra-space component of the higher dimensional gauge field is identified with the Higgs field, called gauge-Higgs unification (GHU) \cite{Manton, Hosotani}, is known to provide a novel solution to the gauge hierarchy problem \cite{Hatanaka}, relying on higher dimensional gauge symmetry, but without invoking supersymmetry.  
Five dimensional (5D) gauge theory is also argued to be related to strongly interacting gauge theory in 4D space-time  \cite{Contino} through AdS/CFT correspondence \cite{Maldacena}, thus leading to a close relationship of GHU with ``little Higgs" scenario. 

In this paper we discuss a scenario where strong CP problem is solved in the framework of higher dimensional gauge theory. In this scenario, the extra-space component of higher dimensional gauge field, say $A_y$ in the case of 5D space-time, behaves as if it were the axion \cite{Peccei, Weinberg} or invisible axion \cite{Kim, Dine} fields in the well-known solutions of the problem in the framework of ordinary 4D space-time.  

The extra-space component $A_y$ is a part of gauge field, not a pseudo Nambu-Goldstone (NG) boson associated with the spontaneous breaking of some global symmetry. However, it still has some similarity to the axion or invisible axion fields. Namely, its coupling with fermion  
field is of pseudo-scalar type from the 4D point of view, being accompanied by $\gamma_{5}$. In addition, the higher dimensional gauge theory possesses a shift symmetry, which is nothing but the higher dimensional local gauge symmetry $A_y \to A_y + \partial_y \lambda$ where $\lambda$ is a $y$-dependent gauge transformation parameter ($y$: extra space coordinate). This mimics the shift symmetry under the global transformation $a \to a + $ constant ($a$: a generic axion field) in those well-known solutions. 

Thus, it is expected that in the higher dimensional gauge theory a solution of the strong CP problem is naturally built-in. 
For concreteness, in this paper we discuss 5D gauge theory with the extra dimension compactified on a circle $S^1$. The coupling of $A_y$ with gluon fields, which plays a crucial role in the solution, is attributed to the ``Chern-Simons" (CS) term, which does not exist in the original 5D theory but is shown to be induced radiatively, as we will see below. The CS term is first derived by a concrete 1-loop calculation and then reconfirmed by utilizing the Fujikawa's method \cite{Fujikawa} to derive the chiral anomaly in 4D space-time, suggesting that the 1-loop calculation of the CS term is the exact result. This may not be surprising, since the relation of parity non-conservation due to the CS term in odd space-time dimensions and the chiral anomaly in even dimensions has been known for some time \cite{Redlich}.   

On the other hand, there appears some novel feature in such obtained CS term, as the inevitable consequence of the fact that the extra space $S^1$ is a non-simply-connected manifold. Namely, $A_y$, or more precisely its Kaluza-Klein (KK) zero mode has a physical meaning as a phase of the Wilson-loop exp$(ie \oint A_{y}dy)$ (in 5D QED, for instance), where the line-integral is along the direction of the extra dimension. This is why the vacuum expectation of $A_y$, which is just a constant gauge field, has a physical importance in GHU. This in turn implies that all physical observables should be periodic in $A_y$. The CS term is not an exception. The loop-induced effective lagrangian obtained by summing up all KK modes in the internal state of the Feynman diagram is no longer linear in $A_y$ as in the ordinary CS term in the de-compactified 5D U(1) gauge theory, but some periodic function of $A_y$. We argue how such novel feature of this scenario causes the modification of the Peccei-Quinn \cite{Peccei} or invisible axion \cite{Kim, Dine} solutions of the strong CP problem. 

In the literature there are works related to the present paper \cite{literature}. In these works, however, the CS term is not derived by concrete calculations of Feynman diagrams, with all KK modes being taken into account in the intermediate state of the loop diagrams. Thus, they do not address the periodicity with respect to the extra-space component $A_y$ nor the non-linearity of the ``Chern-Simons" term with respect to the $A_y$, which are among the main issues of the present paper.

\section{The model} 

In order to discuss the mechanism to solve the strong CP problem, we adopt a simple toy model, i.e. 5-dimensional (5D) SU(3)$_{\rm c} \times$ U(1)$_{\rm X}$ gauge theory, where SU(3)$_{\rm c}$ stands for QCD and U(1)$_{\rm X}$ is some unknown gauge symmetry. As the matter fields, in addition to two flavors of quarks, $u^{\alpha}, \ d^{\alpha}$ ($\alpha = 1,2,3$: color index) we introduce a colored Dirac fermion $\psi^{\alpha}$, which carries U(1)$_{\rm X}$ ``charge" $g$ (the coupling constant of QCD is denoted by $g_{s}$), while aforementioned quarks do not possess the U(1)$_{\rm X}$ interaction. The extra dimension is assumed to be a circle $S^{1}$ of the radius $R$.   

We mainly focus on a part of lagrangian, which is relevant for our purpose:    
\be 
\label{2.1} 
{\cal L} = {\cal L}_{0} + {\cal L}_{\theta} + {\cal L}_{CS}. 
\ee 
${\cal L}_{0}$ is the basic part describing the gauge interactions of matter fermions. The 5D field of $\psi$ is supposed to satisfy an anti-periodic boundary condition $\psi (x^{\mu}, y + 2\pi R) = - \psi(x^{\mu}, y)$ for the convenience of later discussions, while quarks and all gauge fields are supposed to satisfy the periodic boundary conditions:  
\bea 
{\cal L}_{0} &=&  \sum_{n} \overline{\psi^{(n)}} \{ \gamma^{\mu} (i\partial_{\mu} + g_{s} G_{\mu} + gA_{\mu}) + (i\gamma_{5}) 
\left(\frac{n +\frac{1}{2}}{R} + g_{s} G_{y} + gA_{y}\right) - M \} \psi^{(n)} \nonumber \\ 
 &+& \bar{u} \{ \gamma^{\mu} (i\partial_{\mu} + g_{s} G_{\mu}) + (i\gamma_{5}) g_{s}G_{y} - m_{u} \} u + (u \ \to \ d), 
\label{2.2} 
\eea 
where $G_{M} \equiv G^{a}_{M}\frac{\lambda^{a}}{2} = (G_{\mu}, G_{y})$ is 5D gluon field ($\lambda^{a}: {\rm Gell-Mann \ matrices})$ and $A_{M} = (A_{\mu}, A_{y})$ is the 5D U(1)$_{\rm X}$ gauge field. The color indices have been suppressed (for instance, $u$ is understood to be a color triplet: $u = (u^{1}, u^{2}, u^{3})^{t}$). $M, m_{u, d}$ are bulk masses of $\psi$ and quarks, respectively, which are all assumed to be positive unless otherwise stated. 

Though we start from the 5D theory, the lagrangian given in eq.(\ref{2.1}) and eq.(\ref{2.2}) are their effective 4D theories, where the fields $u, d$ and all gauge fields should be understood to denote only their KK zero-modes, while concerning the field $\psi$, all KK modes $\psi^{(n)}$ ($n$: integer) are included, since they all participate in the intermediate state of the loop diagram discussed below. One remark is that, altough we adopt the anti-periodic boundary condition for $\psi$, actually we do  not loose the generality of our discussion by this choice. For instance, if we wish it is possible to move to a base where all fields satisfy the periodic boundary conditions by making a gauge transformation, $\psi \to \psi' = e^{-i \frac{y}{2R}}\psi, \ A_y \to A'_y = A_y - \frac{1}{2gR}$, so that $\psi' (x^{\mu}, y + 2\pi R) = \psi'(x^{\mu}, y)$. We do not discuss here how to remove unnecessary and undesirable interactions, especially those due to $G_{y}$-exchange.

Another piece in eq.(\ref{2.1}), ${\cal L}_{\theta}$ including a parameter $\theta$, reflects the non-trivial vacuum structure of QCD, i.e. the theta vacuum:
\be 
\label{2.3}
{\cal L}_{\theta} = - \theta \frac{g_{s}^{2}}{32\pi^{2}} \epsilon^{\mu \nu \rho \sigma} {\rm Tr} (G_{\mu \nu}G_{\rho \sigma}),
\ee
where $G_{\mu \nu} \equiv G^{a}_{\mu \nu}\frac{\lambda^{a}}{2}$ ($a = 1, 2, \cdots, 8$) is the field strength with respect to the gluon field $G_{\mu}$. To be precise, the $\theta$ should be replaced by $\bar{\theta} = \theta + {\rm arg}({\rm det} M_{q})$, with $M_{q}$ being the mass matrix for quarks, though we simply write $\bar{\theta}$ as $\theta$ in this paper.   

Finally, the ``Chern-Simons (CS) term" ${\cal L}_{CS}$ does not exist in the original action, but is induced through quantum correction as we will see below: 
\be 
\label{2.4} 
{\cal L}_{CS} = \frac{g_{s}^{2}}{32\pi^{2}} f(A_{y}) \epsilon^{\mu \nu \rho \sigma}{\rm Tr} (G_{\mu \nu}G_{\rho \sigma}).
\ee
It is a remarkable feature of this scenario that $A_{y}$ in what we expect from the 5D CS term, $A_{y} \epsilon^{\mu \nu \rho \sigma}{\rm Tr} (G_{\mu \nu}G_{\rho \sigma})$, is inevitably replaced by an odd function $f(A_{y})$ of $A_{y}$ ($f(-A_y) = - f(A_y)$). It is also periodic in $A_y$: $f(A_y + \frac{1}{gR}) = f(A_y)$, which is nothing but the periodicity of the Wilson-loop. As was discussed in the introduction, this property is a natural consequence of the fact that the KK zero mode of $A_{y}$ has a physical meaning as a Wilson-loop (Aharonov-Bohm) phase when the extra space is non-simply-connected space, i.e. $S^{1}$, and therefore all physical observables should be periodic in $A_{y}$.       

When $f(A_y)$ is Taylor-expanded in eq.(\ref{2.4}), the linear term of $A_{y}$, corresponding to the ordinary CS term, is expected to arise generally even in the theories in 4D space-time, where colored matter fields couple with some pseudo-scalar field. In the minimal supersymmetric standard model (MSSM), for instance, the physically remaining pseudo-scalar $A$ decays into two gluons, $A \to gg$, through the effective operator similar to the CS term, $A \epsilon^{\mu \nu \rho \sigma} {\rm Tr}(G_{\mu \nu}G_{\rho \sigma})$, while the Higgs decay $h \to gg$ is handled by an operator $h {\rm Tr}(G^{\mu \nu}G_{\mu \nu})$. The calculations of such effective operators in ordinary 4D space-time have been summarized in the literature \cite{Higgshunters}.  In the next section, we will derive the ``CS term" eq.(\ref{2.4}) in our 5D  model by taking the contributions of all KK modes into account in the intermediate state of the loop-diagram, which inevitably 
leads to the periodic function $f(A_y)$. This function may be understood as the contribution of (the imaginary part of) the Wilson-loop (and its arbitrary powers), which is never expected in the 4D theories.

\section{The ``Chern-Simons" term}   

In this section, we derive the ``Chern-Simons" (CS) term. More precisely, we determine the function $f(A_y)$ in eq.(\ref{2.4}) by two methods, i.e. first by the direct calculation of the relevant 1-loop Feynman diagram and next by utilizing the Fujikawa's method to derive  the chiral anomaly in 4D space-time \cite{Fujikawa}.

\subsection{The derivation by the direct calculation of the 1-loop diagram}

In this subsection, we perform the direct calculation of the 1-loop diagram, contributing to the operator of the form $f(A_{y})\epsilon^{\mu \nu \rho \sigma}{\rm Tr} \{ (\partial_{\mu}G_{\nu} - \partial_{\nu}G_{\mu})(\partial_{\rho}G_{\sigma} - \partial_{\sigma}G_{\rho}) \}$, which is a part of eq.(\ref{2.4}). Its linear term in $A_y$ may be written in the language of the differential form as ${\rm Tr} (AF^{2})$,  
($A, \ F$ are gauge field 1-form and field strength 2-form of a generic gauge group). Note that, although the 5D Chern-Simons form is 
generally written as ${\rm Tr}\{AF^{2} + \frac{1}{2}A^{3}F + \frac{1}{10}A^{5} \}$, when the gauge group contains U(1) factor, as in the case of our model, $A^{3}F, \ A^{5}$ terms just vanish and we do not have to worry about their contributions.  

We just calculate a 1-loop triangle Feynman diagram, contributing to the transition $A_{y} \to GG$. At the first glance, however, it seems that the calculation provides only the term linear in $A_y$. To get $f(A_y)$ with infinite powers of $A_y$, the calculations of arbitrary numbers of $A_y$ field in the external line seem to be necessary. This, however, will be practically impossible.  

We thus need some manipulation. Let us suppose that $A_{y}$ has a vacuum expectation value (VEV) $\bar{A}_{y}$, which really is the case as we will see below, and define a dynamical field $\hat{A}_{y}$ so that $A_{y} = \hat{A}_y + \bar{A}_y$. Then we calculate the triangle diagram to describe $\hat{A}_y \to GG$, where in the propagators of $\psi^{(n)}$ in its internal line we include the contribution of $g \bar{A}_y$ as if it behaves as a mass (of pseudo-scalar type) provided by the VEV. 

On the other hand, if we obtain an operator $\frac{g_{s}^{2}}{32\pi^{2}} f(A_{y})\epsilon^{\mu \nu \rho \sigma}{\rm Tr} \{(\partial_{\mu}G_{\nu} - \partial_{\nu}G_{\mu})(\partial_{\rho}G_{\sigma} - \partial_{\sigma}G_{\rho})\}$, the amplitude for the transition $\hat{A}_{y} \to GG$ is readily obtained by replacing $A_y$ by $\hat{A}_y + \bar{A}_y$ and picking up a term linear in $\hat{A}_y$:  
\be 
\label{4.1} 
\frac{g_{s}^{2}}{32\pi^{2}} f'(\bar{A}_y) \hat{A}_{y} \epsilon^{\mu \nu \rho \sigma}{\rm Tr} \{ (\partial_{\mu}G_{\nu} - \partial_{\nu}G_{\mu})(\partial_{\rho}G_{\sigma} - \partial_{\sigma}G_{\rho})\},   
\ee 
where $f'(\bar{A}_y)$ denotes the first derivative of $f(\bar{A}_y)$.  
Thus, calculating the triangle diagram in a way stated above, it is able to get $f'(\bar{A}_y)$. Then, by replacing it by $f'(A_y)$ and performing an indefinite integral, imposing a condition that $f(A_y)$ should vanish at $A_{y} = 0$ (since $f(A_y)$ should be an odd function of $A_y$ (to preserve the P symmetry (from the viewpoint of the ordinary 4D space-time) of the theory), we can completely determine the function $f(A_y)$, which is also expected to be periodic in $A_y$ as the result of the summation of all KK modes' contributions in the internal line. 

Here is a remark. It should be noticed that the ``CS term" inevitably vanishes for $M = 0$, since both of $M$ and the CS term are ``parity-odd" from the viewpoint of odd-dimensional space-time, such as 5D space-time.\footnote{We will see below that in the specific limit of de-compactification, $R \to \infty$, the CS term survives even for $M = 0$ with $M$-independent coefficient, as is suggested by the index theorem.} This sounds a little strange as the scalar-type mass term $M\bar{\psi}\psi$ is known to be invariant under the ordinary parity transformation in 4D space-time, $P^{(4D)}: \ \psi \to \gamma^{0}\psi$. We know that in 4D space-time the space-inversion $\vec{x} \to - \vec{x}$ caused by this $P^{(4D)}$ is equivalent to the mirror reflection as a discrete transformation, since they are mutually related by a space rotation.  In the 5D space-time, on the other hand, the $P^{(4D)}$ transformation causes $(\vec{x}, y) \to -(\vec{x}, y)$. This is no longer a discrete transformation but one of the space rotations, which is an automatic symmetry of the theory and is not useful for investigating the property of the CS term. If we adopt an alternative discrete transformation, say $P^{(5D)}: \ \psi \to \gamma^{0}\gamma_{5}\psi$, causing $(\vec{x}, y) \to (- \vec{x}, y)$, $\bar{\psi}\psi$ is easily known to be odd under this transformation, while $\bar{\psi}(i\gamma_{5})\psi$ is invariant instead. In such a sense, $M$ is a unique parity-odd parameter in the lagrangian for $\psi$, while the CS term itself also becomes odd under the $P^{(5D)}$. Therefore, the coefficient of the CS term should be accompanied by some odd function of $M$ and vanishes for $M = 0$.     

\begin{figure}[bt] 
\includegraphics[bb=0 0 100 100]{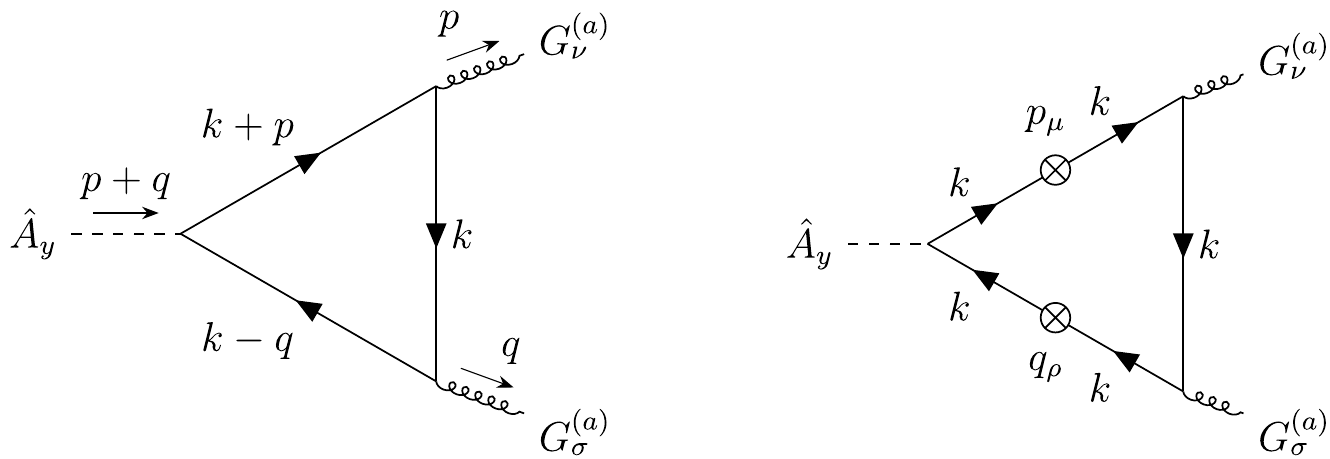} 
\caption{The Feynman diagram generating the ``Chern-Simons" term} 
\label{fig:1} 
\end{figure}

Now we calculate the triangle diagram for the process $\hat{A}_y(p + q) \to G^{(a)}_{\nu}(p) + G^{(a)}_{\sigma}(q)$ shown in the left-hand side of Fig.\ref{fig:1}, with $p, q$ being 4-momenta carried by $G^{(a)}_{\nu}$ and $G^{(a)}_{\sigma}$, respectively. Actually, what we are interested in is the operator with derivatives on the gluon fields, $\epsilon^{\mu \nu \rho \sigma}{\rm Tr} \{ (\partial_{\mu}G_{\nu})(\partial_{\rho}G_{\sigma})\}$. Since $\partial_{\mu}$ on $G^{(a)}_{\nu}(p)$ is equivalent to the multiplication of $G^{(a)}_{\nu}(p)$ by $i p_{\mu}$ in the momentum space, for instance, we pick up the term, which is linear in both of $p_{\mu}$ and $q_{\rho}$, by taking derivatives of the triangle diagram with respect to $p_{\mu}$ and $q_{\rho}$ and then setting $p = q = 0$. Thus, effectively what we calculate is a pentagon diagram with 5 internal propagators, as is shown in the right-hand side of Fig.\ref{fig:1}. The Wilson coefficient of the operator $\hat{A}_{y}(\partial_{\mu}G^{(a)}_{\nu}) (\partial_{\rho}G^{(a)}_{\sigma})$ is known to be given as (by noting ${\rm Tr}(T^{(a)}T^{(b)}) = \frac{1}{2}\delta^{ab}$)  
\bea 
&& -i \frac{gg_{s}^{2}}{2} \sum_{n}\int \frac{d^{4}k}{(2\pi)^{4}} \frac{1}{[k^{2} -(\frac{n +\frac{1}{2}}{R} + g \bar{A}_{y})^{2} - M^{2}]^{5}}\nonumber \\ 
&& \cdot {\rm Tr}\{ i\gamma_{5} [\Slash{k} + i\gamma_{5}(\frac{n +\frac{1}{2}}{R} + g \bar{A}_{y}) + M] \gamma^{\mu} [\Slash{k} + i\gamma_{5}(\frac{n +\frac{1}{2}}{R} + g \bar{A}_{y}) + M] \gamma^{\nu} \nonumber \\ 
&& \cdot [\Slash{k} + i\gamma_{5}(\frac{n +\frac{1}{2}}{R} + g \bar{A}_{y}) + M] \gamma^{\sigma}[\Slash{k} + i\gamma_{5}(\frac{n +\frac{1}{2}}{R} + g \bar{A}_{y}) + M] \gamma^{\rho} \nonumber \\ 
&& \cdot [\Slash{k} + i\gamma_{5}(\frac{n +\frac{1}{2}}{R} + g \bar{A}_{y}) + M] \}.   
\label{4.3}    
\eea 

By picking up terms with odd powers of $i\gamma_{5}$ inside the ``Tr", the numerator of the integrand is simplified (after some lengthy but straightforward arithmetic) into 
\be 
\label{4.4} 
\{ (k^{2})^{2}M-2k^{2}M^{3}+M^{5}-2M(k^{2}-M^{2})(\frac{n +\frac{1}{2}}{R} + g \bar{A}_{y})^{2}+M(\frac{n +\frac{1}{2}}{R} + g \bar{A}_{y})^{4} \}{\rm Tr}(i\gamma_{5} \gamma^{\mu}\gamma^{\nu}\gamma^{\sigma}\gamma^{\rho}), 
\ee 
where we have used the properties (for all different $\mu, \ \nu, \rho, \sigma$ and under the integral $\int d^{4}k$)
\be 
\label{4.5} 
\Slash{k}\gamma^{\mu}\Slash{k} = - \frac{k^{2}}{2}\gamma^{\mu}, \ \Slash{k}\gamma^{\mu}\gamma^{\nu}\Slash{k} = 0, \ 
\Slash{k}\gamma^{\mu}\gamma^{\nu} \gamma^{\rho}\Slash{k} = \frac{k^{2}}{2}\gamma^{\mu}\gamma^{\nu}\gamma^{\rho}, \ 
\Slash{k}\gamma^{\mu}\gamma^{\nu} \gamma^{\rho}\gamma^{\sigma}\Slash{k} = - k^{2}\gamma^{\mu}\gamma^{\nu}\gamma^{\rho}\gamma^{\sigma},  
\ee 
and ${\rm Tr}(i\gamma_{5}\Slash{k} \cdots \Slash{k}) = - k^{2}{\rm Tr}(i\gamma_{5} \cdots)$. Note that ${\rm Tr}(i\gamma_{5} \gamma^{\mu}\gamma^{\nu}\gamma^{\sigma}\gamma^{\rho})$ survives only when all of $\gamma^{\mu}, \gamma^{\nu}, \gamma^{\sigma}, \gamma^{\rho}$ are different. 

As ${\rm Tr} (i\gamma_{5}\gamma^{\mu}\gamma^{\nu}\gamma^{\sigma}\gamma^{\rho}) = 4 \epsilon^{\mu \nu \sigma \rho}$, we have obtained 
an operator $\epsilon^{\mu \nu \sigma \rho}\hat{A}_{y}(\partial_{\mu}G^{(a)}_{\nu}) (\partial_{\rho}G^{(a)}_{\sigma}) = 
-\frac{1}{4}\epsilon^{\mu \nu \rho \sigma}\hat{A}_{y}(\partial_{\mu}G^{(a)}_{\nu} - \partial_{\nu}G^{(a)}_{\mu}) (\partial_{\rho}G^{(a)}_{\sigma} - \partial_{\sigma}G^{(a)}_{\rho})$, which is nothing but the one in eq.(\ref{4.1}), whose Wilson coefficient is given by 
\bea 
&&\frac{g_{s}^{2}}{32\pi^{2}} f'(\bar{A}_y)  \nonumber \\ 
&=& i gg_{s}^{2}M \sum_{n}\int \frac{d^{4}k}{(2\pi)^{4}} \frac{(k^{2})^{2}-2k^{2}M^{2}+M^{4}-2(k^{2}-M^{2})(\frac{n +\frac{1}{2}}{R} + g \bar{A}_{y})^{2}+(\frac{n +\frac{1}{2}}{R} + g \bar{A}_{y})^{4}}{[k^{2} -(\frac{n +\frac{1}{2}}{R} + g \bar{A}_{y})^{2} - M^{2}]^{5}} \nonumber \\ 
&=& i gg_{s}^{2}M \sum_{n}\int \frac{d^{4}k}{(2\pi)^{4}} \frac{1}{[k^{2} -(\frac{n +\frac{1}{2}}{R} + g \bar{A}_{y})^{2} - M^{2}]^{3}}. 
\label{4.6} 
\eea 
By use of 
\be 
\label{4.7}
\int \frac{d^{4}k}{(2\pi)^{4}} \frac{1}{(k^{2}-\Delta)^{3}} = -\frac{i}{(4\pi)^{2}}\frac{1}{2}\frac{1}{\Delta},  
\ee 
we find  
\be 
\label{4.8} 
f'(\bar{A}_y) = gM \sum_{n} \frac{1}{(\frac{n +\frac{1}{2}}{R} + g \bar{A}_{y})^{2} + M^{2}}. 
\ee 

The KK mode sum is performed utilizing a mathematical formula, 
\be 
\label{4.9} 
\sum_{n} \frac{1}{(\frac{n +\frac{1}{2}}{R} + g \bar{A}_{y})^{2} + M^{2}} = \frac{\pi R}{M} \frac{\sinh (2\pi RM)}{\cosh (2\pi RM) + \cos (2\pi gR\bar{A}_{y})}, 
\ee 
to get 
\be 
\label{4.10} 
f'(\bar{A}_y)  = g (\pi R) \frac{\sinh (2\pi RM)}{\cosh (2\pi RM) + \cos (2\pi gR\bar{A}_{y})}. 
\ee 

The final task is to perform the indefinite integral of $f'(A_y)$, obtained by replacing $\bar{A}_y$ by $A_y$ in eq.(\ref{4.10}): 
\be 
\label{4.13} 
f(A_y) = g (\pi R) \ \int \ \frac{\sinh (2\pi RM)}{\cosh (2\pi RM) + \cos (2\pi gRA_{y})} \ dA_{y}. 
\ee 
Unfortunately, it is hard to perform this integral. We, however, succeed in writing the $f(A_y)$ in the form of KK mode-sum. Namely, the indefinite integral of $\frac{1}{(\frac{n +\frac{1}{2}}{R} + g A_{y})^{2} + M^{2}}$, which is obtained by replacing $\bar{A}_y$ by $A_y$ in eq.(\ref{4.8}), 
is able to be performed by the change of integral variable $A_y \ \to \ \alpha_{n}$, so that $\frac{n +\frac{1}{2}}{R} + gA_y = M \cot \alpha_{n}$, i.e. $\alpha_{n} = \cot^{-1}\left( \frac{\frac{n +\frac{1}{2}}{R} + gA_y}{M} \right) = \tan^{-1}\left( \frac{M}{\frac{n +\frac{1}{2}}{R} + gA_y} \right)$: 
\be 
\label{4.14} 
\int \frac{1}{(\frac{n +\frac{1}{2}}{R} + g A_{y})^{2} + M^{2}} \ dA_{y} = - \frac{1}{gM} \int \ d\alpha_{n} = - \frac{1}{gM}\alpha_{n}, 
\ee 
where an additional constant of integral is not allowed, since its presence contradicts with the condition that $f(A_y)$ is an odd function of $A_y$. 
Hence, from eq.(\ref{4.8}) $f(A_y)$ is found to be  
\be 
\label{4.15}
f(A_y) =  g M \sum_{n} \int \frac{1}{(\frac{n +\frac{1}{2}}{R} + g A_{y})^{2} + M^{2}} \ dA_y = - \sum_{n} \alpha_{n} = - \sum_{n} \tan^{-1}\left( \frac{M}{\frac{n +\frac{1}{2}}{R} + gA_y}\right). 
\ee 
$f(A_y)$ is an odd-function of $A_y$, as we easily see by replacing $A_y$ by $- A_y$ and $n + \frac{1}{2}$ by $-n - \frac{1}{2}$ in eq.(\ref{4.15}), in particular $f(0) = 0$. Note that if we adopted a periodic boundary condition for 5D field $\psi$, we would have obtained $f(A_y) =  - \sum_{n} \tan^{-1}(\frac{M}{\frac{n}{R} + gA_y})$, which is discontinuous at the origin $A_y = 0$ because of the contribution of the KK zero-mode: $\lim_{A_y \to \pm 0} f(A_y) = \mp \frac{\pi}{2}$. This is the reason why we chose the anti-periodic boundary condition for $\psi$.       

Thus we finally have succeeded in getting the ``Chern-Simons" term, though it is not linear in $A_y$, but a periodic function of $A_y$: 
\bea 
{\cal L}_{CS} &=& \frac{g_{s}^{2}}{32\pi^{2}} f(A_{y}) \epsilon^{\mu \nu \rho \sigma}{\rm Tr} (G_{\mu \nu}G_{\rho \sigma}) \nonumber \\ 
&=& - \frac{g_{s}^{2}}{32\pi^{2}}\sum_{n} \tan^{-1}\left( \frac{M}{\frac{n +\frac{1}{2}}{R} + gA_y} \right) \ \cdot \ \epsilon^{\mu \nu \rho \sigma}{\rm Tr} (G_{\mu \nu}G_{\rho \sigma}).   
\label{4.16} 
\eea 
As we anticipated, this CS term vanishes for $M = 0$.

Here appears a question whether the KK mode-sum in eq.(\ref{4.16}) is convergent, i.e. whether the CS term is free from UV-divergence. When we consider the UV-divergence, it is enough to think about what happens in the de-compactified limit, where the KK mode-sum is reduced to the integral $\int dk_y$ ($k_y$: the 5-th component of the 5D momentum).  In the 5D space-time the operator linear in $A_y$, $g g_{s}^{2}A_y \epsilon^{\mu \nu \rho \sigma}{\rm Tr} (G_{\mu \nu}G_{\rho \sigma})$ in the CS term has mass dimension 5 (with gauge couplings being included in the gauge fields, $g A_y$ for instance). So at the first glance it seems that its Wilson coefficient suffers from a logarithmic UV-divergence. Actually, however, as we have already seen, the coefficient should be accompanied by an overall factor $M$, as it vanishes for $M = 0$, and the momentum integral behaves at the UV-region as $\int d^{5}k/(k^{2})^{3}$, which is convergent.  
Thus, the predicted CS term should be free from UV-divergences. This argument is supported by rewriting eq.(\ref{4.16}) as follows, 
\be 
\label{4.16'} 
{\cal L}_{CS} = \frac{g_{s}^{2}}{32\pi^{2}}\sum_{n=1}^{\infty} \tan^{-1}\left( \frac{8gR^{2}MA_y}{(2n -1)^{2} +4R^{2}[M^{2}-(gA_y)^{2}]} \right) \ \cdot \ \epsilon^{\mu \nu \rho \sigma}{\rm Tr} (G_{\mu \nu}G_{\rho \sigma}), 
\ee 
which should be convergent, since for large $n$, the mode-sum behaves as $\sim \sum_{n} (1/(2n -1)^{2}) \simeq (1/4) \sum_{n} (1/n^{2}) $. 

Interestingly, at the limit of de-compactification of 5D space-time, $R \to \infty$, eq.(\ref{4.16}) is found to be reduced to the ordinary 5D Chern-Simons term. First we note that in the Taylor expansion of $f(A_y)$ around its origin, from eq.(\ref{4.10}) the first derivative is reduced to 
\be 
\label{4.17} 
\lim_{R \to \infty}\frac{f'(0)}{2\pi R}  = \frac{g}{2},  
\ee
where the factor $1/(2\pi R)$ is added since in general the lagrangians for the KK zero-mode in 4D and 5D space-time, say ${\cal L}_{{\rm 4D}}, \ {\cal L}_{{\rm 5D}}$, are related by ${\cal L}_{{\rm 4D}} = (2\pi R) {\cal L}_{{\rm 5D}}$. We also easily realize from eq.(\ref{4.10}) that the higher derivatives $f^{(2n+1)}(0) \ (n = 1,2, \cdots)$, denoting $2n+1$-th derivatives, all vanish in the limit of $RM \to \infty$ because of the exponential suppression factor $e^{-2\pi RM}$ ($f^{(2n)}(0) \ (n = 1,2, \cdots)$ identically vanish, as $f(A_y)$ is an odd function).  Thus, in the limit of de-compactification, only the ordinary 5D Chern-Simons term is left:  
\be 
\label{4.18} 
{\cal L}_{{\rm 5D-CS}} = \frac{gg_{s}^{2}}{64\pi^{2}}A_{y} \epsilon^{\mu \nu \rho \sigma}{\rm Tr} (G_{\mu \nu}G_{\rho \sigma}),  
\ee 
whose coefficient is independent of $M$ and also UV-finite, as we expected. To be precise, although we have assumed that $M$ is positive, it is easy to confirm that in general eq.(\ref{4.18}) is accompanied by $\frac{M}{|M|} = {\rm sgn} (M)$. Let us note that the combinations $gA_y, \ g_{s}G_{\mu}$ can be readily re-interpreted as the corresponding 5D quantities. 

Thus, in this specific limit, the periodicity in $A_y$ is lost. This may be a natural result, since in the limit of de-compactification, the non-local operator, Wilson-loop, loses its physical importance and only the local ``gauge invariant" (in the sense that gauge transformation yields a total derivative) operator is allowed.

\subsection{The derivation based on the Fujikawa's method}   

In the previous subsection, the ``Chern-Simons" term eq.(\ref{4.16}) was derived by the explicit calculation of 1-loop Feynman diagram. It will be natural to ask 
whether it is modified by the quantum corrections at higher loops, such as two-loop correction. From this point of view, it may be suggestive to 
recall that the 1-loop contribution to the chiral anomaly is exact, i.e. there do not exist corrections from higher loops. Since the operator eq.(\ref{4.16}) has much similarity to the chiral anomaly, it may be reasonable to expect that eq.(\ref{4.16}) is actually an exact result. In this section, we would like to demonstrate that this really is the case, utilizing the Fujikawa's interpretation of the chiral anomaly \cite{Fujikawa}: it is solely due to the change of measure for the chiral transformation in the language of fermionic path-integral. This may be a natural consequence, since the relation of parity non-conservation due to the CS term in odd space-time dimensions and the chiral anomaly in even dimensions has been known for some time \cite{Redlich}.  

We start from the lagrangian for the fermionic field $\psi$, which include all KK modes $n$ (refer to eq.(\ref{2.2})):  
\be 
\label{5.1} 
{\cal L}_{\psi} = \sum_{n} \overline{\psi^{(n)}} \{ \gamma^{\mu} (i\partial_{\mu} + g_{s} G_{\mu}) + (i\gamma_{5}) (\frac{n +\frac{1}{2}}{R} + gA_{y}) - M \} \psi^{(n)}, 
\ee  
where the parts including the interactions due to $A_{\mu}$ and $G_y$ have been ignored, since they are unnecessary for our argument below.

In the discussion on the CS term based on the symmetry under $P^{(5D)}$ we learned that ${\cal L}_{CS}$ should vanish exactly for $M = 0$. However, even for non-vanishing $M$ we realize that the  scalar-type mass term $- M \overline{\psi^{(n)}}\psi^{(n)}$ may be eliminated by performing a suitable chiral transformation of $\psi^{(n)}$. First, note that the mass operator for $\psi^{(n)}$ as the whole can be written as  
\be 
\label{5.2}
(i\gamma_{5}) (\frac{n +\frac{1}{2}}{R} + gA_{y}) - M = (i\gamma_{5}) \sqrt{(\frac{n +\frac{1}{2}}{R} + gA_{y})^{2} + M^{2}} \cdot e^{2i\gamma_{5}\kappa_{n}}, 
\ee 
where $\tan (2\kappa_{n}) = M/(\frac{n +\frac{1}{2}}{R} + gA_{y})$. Then, it is easy to see that by performing a chiral transformation 
\be 
\label{5.3} 
\psi^{(n)} \ \to \ \psi'^{(n)} = e^{i\gamma_{5}\kappa_{n}} \psi^{(n)}, 
\ee 
the scalar-type mass term is eliminated. Then we have to conclude that the quantum corrections due to the exchange of $\psi'^{(n)}$ never yield ${\cal L}_{CS}$ in all orders of perturbation, since the transformed lagrangian preserves $P^{(5D)}$ symmetry. We, however, have to be a little careful. Namely, the chiral transformation leads to an additional term in the lagrangian, originating from the change of the measure in the path integral of $\psi^{(n)}$, i.e. from the chiral anomaly \cite{Fujikawa}: 
\be 
\label{5.4} 
- \frac{g_{s}^{2}}{32\pi^{2}}(2\kappa_{n}) \epsilon^{\mu \nu \rho \sigma} {\rm Tr}(G_{\mu \nu}G_{\rho \sigma}). 
\ee 
Summing up the contributions of all KK modes, we find 
\bea 
{\cal L}_{{\rm anomaly}} &=& - \sum_{n}\frac{g_{s}^{2}}{32\pi^{2}}(2\kappa_{n}) \epsilon^{\mu \nu \rho \sigma}{\rm Tr} (G_{\mu \nu}G_{\rho \sigma})  \nonumber \\ 
&=& - \frac{g_{s}^{2}}{32\pi^{2}}\sum_{n} \tan^{-1}\left( \frac{M}{\frac{n +\frac{1}{2}}{R} + gA_y} \right) \ \cdot \ \epsilon^{\mu \nu \rho \sigma} {\rm Tr} (G_{\mu \nu}G_{\rho \sigma}),  
\label{5.5} 
\eea 
which is exactly the same as eq.(\ref{4.16}). 

This result is exact, in the sense that it is correct in all orders of perturbation. We thus conclude that the result eq.(\ref{4.16}) is, say, 1-loop exact. One may wonder whether the bulk masses of quarks, $m_{u, d}$, also contribute to the CS term and modify the result. Actually, since quarks do not have couplings with $A_y$, even if chiral transformation for the quark fields, similar to the one for $\psi$, is made, they do not produce the CS term. It is interesting to note that this in turn means the CS term is free from UV-divergence, even though the theory itself, being higher dimensional gauge theory, is non-renormalizable.

\section{The mechanism to solve the strong CP problem}  

A remarkable feature of our model is that it has some similarity to the Peccei-Quinn (PQ) model \cite{Peccei, Weinberg} or the models of invisible axion \cite{Kim, Dine}. The essence of the mechanism to solve the strong CP problem in these models would be to replace the parameter $\theta$ (or more precisely $\bar{\theta}$) by a dynamical variable by introducing axion field $a$, which is a NG boson due to the spontaneous breaking of global U(1)$_{PQ}$ symmetry of the theory. Under the U(1)$_{PQ}$ transformation $a$ is shifted as $a \ \to \ a +c$ ($c$: constant): the theory has a shift symmetry. The U(1)$_{PQ}$ symmetry is a sort of chiral symmetry, and since the chiral symmetry has an anomaly $\partial_{\mu}j^{\mu}_{5} = - \frac{g_{s}^{2}}{32\pi^{2}} \epsilon^{\mu \nu \rho \sigma}G^{a}_{\mu \nu}G^{a}_{\rho \sigma}$, in the lagrangian there arises the following interaction term of the $a$ with gluon fields: 
\be 
\label{6.1} 
- \frac{g_{s}^{2}}{32\pi^{2}} \frac{a}{V}\epsilon^{\mu \nu \rho \sigma}{\rm Tr} (G_{\mu \nu}G_{\rho \sigma}), 
\ee 
where $V$ is the typical mass scale relevant for the spontaneous breaking of the global symmetry. 
Now the parameter $\theta$ effectively has been replaced by a dynamical variable $\theta + \frac{a}{V}$. Redefining the axion field as $\frac{\hat{a}}{V} \equiv \theta + \frac{a}{V}$ ($\hat{a} = a - (- V \theta)$), the potential for $\hat{a}$ due to QCD interaction is shown to have a minimum at $\hat{a} = 0$. Thus, the P and CP violating parameter $\theta$ is removed from the theory and the strong CP problem is solved. (In this section, as we discuss the 4D low-energy effective theory, the P and CP transformations are defined in the ordinary manner: e.g. P: $u \to \gamma^{0}u$, etc. Hence the quark masses $m_{u, d}$ are P and CP preserving, as usual.)   

We realize that our model has some similarity to these models. Comparing eq.(\ref{2.4}) and eq.(\ref{6.1}), we find that the role of the axion field $a$ is played by $A_{y}$ in our model, though in our case, eq.(\ref{2.4}), $f(A_y)$ is not linear in $A_y$ but a periodic function of $A_y$. We notice that $A_{y}$ also has a shift symmetry, though it is not a NG boson, which is associated with the local gauge symmetry, not a global symmetry: the symmetry under the local gauge transformation $A_{y} \ \to \ A_{y} + \partial_{y} \lambda$ ($\lambda$: $y$-dependent gauge transformation parameter). 

We are now ready to discuss our mechanism to solve the strong CP problem. First, let us focus on the combination of the theta term and the CS term, eq.(\ref{2.3}) and eq.(\ref{2.4}), 
\be 
\label{6.2}  
{\cal L}_{\theta} + {\cal L}_{CS}= \frac{g_{s}^{2}}{32\pi^{2}} [- \theta + f(A_y)] \epsilon^{\mu \nu \rho \sigma} {\rm Tr} (G_{\mu \nu}G_{\rho \sigma}). 
\ee 
We assume that the VEV $\bar{A}_y$ is tuned as 
\be 
\label{6.2'} 
f(\bar{A}_y) = \theta,  
\ee 
so that eq.(\ref{6.2}) vanishes for the VEV. As is seen below, this assumption is justified by showing that in the low-energy effective theory, the potential for $\hat{A}_y = A_y - \bar{A}_y$ and pion fields, both having odd intrinsic parities being pseudo-scalars, naturally starts from the quadratic terms of these fields, since our model is P and CP conserving, and therefore its minimum is expected to be at $\hat{A}_y = \pi^{0} = 0$.

Then eq.(\ref{6.2}) is rewritten by use of ${\cal F}(\hat{A}_y) \equiv f(\bar{A}_y + \hat{A}_y) - f(\bar{A}_y) = f'(\bar{A}_y)\hat{A}_y + \cdots \ ({\cal F}(0) = 0)$, as  
\be 
\label{6.3}  
{\cal L}_{\theta} + {\cal L}_{CS}= \frac{g_{s}^{2}}{32\pi^{2}} {\cal F}(\hat{A}_y) \epsilon^{\mu \nu \rho \sigma} {\rm Tr} (G_{\mu \nu}G_{\rho \sigma}). 
\ee 
In this way, the lagrangian, now ignoring the field $\psi$ (and assuming that the unnecessary field $G_{y}$ is somehow removed from the theory), reads as 
\bea 
{\cal L} &\supset& \frac{g_{s}^{2}}{32\pi^{2}} {\cal F}(\hat{A}_y) \epsilon^{\mu \nu \rho \sigma} {\rm Tr}(G_{\mu \nu}G_{\rho \sigma})  \nonumber \\ 
   &+& \bar{u} \{ \gamma^{\mu} (i\partial_{\mu} + g_{s} G_{\mu}) - m_{u} \} u + (u \ \to \ d). 
\label{6.4} 
\eea 

Now we invoke the chiral transformation of quark fields in order to eliminate the first line of eq.(\ref{6.4}), proportional to ${\cal F}(\hat{A}_y)$. The purpose of this procedure is to get the potential of $\hat{A}_{y}$ (and the neutral pion field $\pi^{0}$) expected in the low-energy effective theory. We perform chiral transformations of the quark fields $u, \ d$: 
\be 
\label{6.5}
u \ \to e^{i\gamma_{5}\alpha_{u}}u, \ \ \ d \ \to e^{i\gamma_{5}\alpha_{d}}d.
\ee 
Then the quark mass term in eq.(\ref{6.4}) is modified into $- m_{u}\bar{u} e^{2i\gamma_{5}\alpha_{u}}u + (u \ \to \ d)$. In addition, because of the chiral anomaly, the lagrangian acquires an additional term given by 
\be 
\label{6.6}
\frac{g_{s}^{2} (\alpha_{u} + \alpha_{d})}{16\pi^{2}} \epsilon^{\mu \nu \rho \sigma} {\rm Tr}(G_{\mu \nu}G_{\rho \sigma}). 
\ee  
Now by comparing eq.(\ref{6.4}) and eq.(\ref{6.6}), a chiral transformation, satisfying 
\be 
\label{6.7}
\alpha_{u} + \alpha_{d} = - \frac{1}{2}{\cal F}(\hat{A}_y) \ \ \to \ \  \alpha_{u} = - a_{u}\frac{1}{2}{\cal F}(\hat{A}_y), \ \alpha_{d} = - a_{d}\frac{1}{2}{\cal F}(\hat{A}_y),
\ee 
is known to eliminate the first line of eq.(\ref{6.4}). Here $a_{u, d}$ are constant coefficients satisfying $a_{u} + a_{d} = 1$. 

The resulting low-energy effective lagrangian for $\hat{A}_{y}$ and $\pi^{0}$ is obtained by making the following replacements \cite{Weinberg2}: 
\bea 
&&\bar{u}u \ \ \to \ \ -v \cos (\frac{\pi_{0}}{F_{\pi}}), \ \ \bar{d}d \ \ \to \ \ -v \cos (\frac{\pi_{0}}{F_{\pi}}),  
\label{6.8a} \\ 
&&\bar{u}\gamma_{5}u \ \ \to \ \ -iv \sin (\frac{\pi_{0}}{F_{\pi}}), \ \ \bar{d}\gamma_{5}d \ \ \to \ \ iv \sin (\frac{\pi_{0}}{F_{\pi}}),   
\label{6.8b} \\ 
&&\bar{u}\gamma^{\mu}\gamma_{5}u \ \ \to \ \ \frac{1}{2}F_{\pi}\partial^{\mu}\pi^{0}, \ \ \bar{d}\gamma^{\mu}\gamma_{5}d \ \ \to \ \ 
-\frac{1}{2}F_{\pi}\partial^{\mu}\pi^{0},   
\label{6.8c} 
\eea 
where $v = |\langle \bar{u}u \rangle| = |\langle \bar{d}d \rangle|$ and $F_{\pi}$ is the decay constant of the pions. 

The relevant part of such obtained low energy effective lagrangian is given as 
\bea 
{\cal L}_{eff} &=& \frac{1}{2}(\partial_{\mu}\pi^{0})(\partial^{\mu}\pi^{0}) + \frac{1}{2}(\partial_{\mu}\hat{A}_{y})(\partial^{\mu}\hat{A}_{y}) 
+ \frac{a_{u}-a_{d}}{4}F_{\pi}(\partial_{\mu}\pi^{0})(\partial^{\mu}{\cal F}(\hat{A}_{y}))  
\nonumber  \\ 
&+& m_{u} \ v \cos \left[ \frac{\pi^{0}}{F_{\pi}} - a_{u}{\cal F}(\hat{A}_{y}) \right] + m_{d} \ v \cos \left[ \frac{\pi^{0}}{F_{\pi}} + a_{d}{\cal F}(\hat{A}_{y}) \right].   
\label{6.9} 
\eea 
To simplify the analysis we use the freedom of taking a choice of $a_{u} = a_{d} = \frac{1}{2}$.  Then the mixing in the kinetic term between $\hat{A}_{y}$ and $\pi^{0}$ disappears and these fields become correctly normalized fields. Thus, we finally get the effective potential for these two pseudo-scalar fields,         
\be 
V_{eff} = - m_{u} \ v \cos \left[ \frac{\pi^{0}}{F_{\pi}} - \frac{1}{2}{\cal F}(\hat{A}_{y}) \right] - m_{d} \ v \cos \left[ \frac{\pi^{0}}{F_{\pi}} + \frac{1}{2}{\cal F}(\hat{A}_{y}) \right]. 
\label{6.10} 
\ee 

The quadratic part of $\pi^{0}$ and $\hat{A}_{y}$ in eq.(\ref{6.10}) is easily calculated by noting ${\cal F}(\hat{A}_y) = f'(\bar{A}_y)\hat{A}_y + \cdots$, and the mass-squared matrix $M_{ps}^{2}$ in the basis of $\pi^{0}$ and $\hat{A}_{y}$ is given as 
\be 
\label{6.11}
M_{ps}^{2} = 
\begin{pmatrix}
\frac{v}{F_{\pi}^{2}}(m_{u} + m_{d}) &  - \frac{v}{F_{\pi}F_{A_y}}(m_{u} - m_{d})  \cr 
- \frac{v}{F_{\pi}F_{A_y}}(m_{u} - m_{d}) & \frac{v}{F_{A_y}^{2}} (m_{u} + m_{d}) \cr 
\end{pmatrix},  
\ee 
where $F_{A_y} \equiv \frac{2}{f'(\bar{A}_y)}$. 
Obviously Tr $M_{ps}^{2} > 0$ and also det $M_{ps}^{2} = 4 \frac{v^{2}}{F_{\pi}^{2}F_{A_y}^{2}}m_{u}m_{d} > 0$. Thus the two real eigenvalues of this matrix are both positive, and we conclude that the minimum of the potential is at $\hat{A}_y = \pi^{0} = 0$, as we expected. Hence, the strong CP problem is resolved. 

Under a reasonable assumption (see the discussions below) of $F_{A_y} \gg F_{\pi}$, one eigenvalue of $M_{ps}^{2}$ gives the well-known formula for the pion mass: $m_{\pi}^{2} = \frac{v}{F_{\pi}^{2}}(m_{u} + m_{d})$. Having the det $M_{ps}^{2}$, another eigenvalue is readily obtained, which is nothing but the mass-squared of the axion-like particle, whose dominant component is $\hat{A}_y$:  
\be 
\label{6.12} 
m_{A_y}^{2} = \frac{v}{F_{A_y}^{2}}\frac{4m_{u}m_{d}}{m_{u} + m_{d}} = \left( \frac{F_{\pi}}{F_{A_y}} \right)^{2}\frac{4m_{u}m_{d}}{(m_{u} + m_{d})^{2}} m_{\pi}^{2}.  
\ee 

The size of the mass scale $F_{A_y}$ is limited by astrophysical and cosmological arguments \cite{Weinberg2, PDG}. As a rough estimation, the observation of the supernova SN1987A 
puts an lower bound $F_{A_y} > 10^{10}$ GeV, while cosmological arguments suggest $F_{A_y} < 10^{12}$ GeV. Thus rather narrow window is left for the allowed region: $10^{10} {\rm GeV} < F_{A_y} < 10^{12} {\rm GeV}$. We now investigate how the parameters of our model, such as $M_{c}, M, g$, are constrained by this condition. Here $M_{c} \equiv 1/R$ is the ``compactification mass scale" of the extra dimension. From eq.(\ref{4.10}), depending on the relative size of $M$ compared to $M_{c}$, $F_{A_y}$ behaves as 
\bea 
F_{A_y} &=& \frac{2}{f'(\bar{A}_y)} = \frac{2M_{c}}{\pi g} \frac{\cosh (2\pi \frac{M}{M_{c}}) + \cos (2\pi g\frac{\bar{A}_{y}}{M_{c}})}{\sinh (2\pi \frac{M}{M_{c}})} \nonumber \\ 
&\simeq&   
\begin{cases}  
\frac{2}{\pi} \left( \frac{M_{c}}{g} \right) \sim \frac{M_{c}}{g}  & ({\rm for} \ M \gg M_{c}) \\ 
\frac{2}{\pi^{2}} \frac{M_{c}^{2}}{gM} \cos^{2}(\pi g \frac{\bar{A}_y}{M_{c}}) \sim \frac{M_{c}^{2}}{gM} & ({\rm for} \ M \ll M_{c})   
\end{cases}, 
\label{6.13}
\eea 
unless $g \frac{\bar{A}_y}{M_{c}} \simeq \frac{1}{2}$ accidentally. 

For the case of $M \gg M_{c}$, first let us suppose $g$ is of ${\cal O}(1)$, to get a rough idea. Then, $M_{c}$ should be of ${\cal O}(10^{10-12})$ GeV, say the intermediate mass scale between the weak scale $M_{W}$ and the Planck scale $M_{pl}$. However, if this model is embedded into the electro-weak gauge-Higgs unification model (on a flat 5D space-time) \cite{Kubo}, the compactification scale $M_{c}$ is expected to be not so far from the weak scale, say $M_{c} \sim$ 1 to 10 TeV. Then, extremely small gauge coupling $g$ is needed. For instance in the case of $M_{c} \sim 10$ TeV, $g \sim 10^{-7}$ is required in order to realize $F_{A_y} \sim 10^{11}$ GeV. 

Another possibility is to invoke the property that $A_y$ decouples from the low energy theory at the limit of $M \to 0$, since in this limit the CS term itself disappears. In fact, as is seen in eq.(\ref{6.13}), $F_{A_y} \to \infty$ in this limit. So even for $g ={\cal O}(1)$, the constraint on $M_{c}$ is considerably relaxed. For instance, assuming $M_{c} \sim 10$ TeV, $M \sim 1$ MeV realizes $F_{A_y} \sim 10^{11}$ GeV. Note that even for such small $M$, the mass of the colored fermion $\psi^{(0)}$ is estimated to be $\sqrt{(\frac{1}{2R}+g\bar{A}_y)^{2} + M^{2}} \sim \frac{M_{c}}{2} \sim 10$ TeV, and therefore safely evades experimental search.

\section{Summary and discussion} 

In this paper we discussed a natural scenario to solve the strong CP problem in the framework of the higher dimensional gauge theory. 
It has a similarity to the well-known Peccei-Quinn \cite{Peccei} or invisible axion \cite{Kim, Dine} solutions of the strong CP problem. Axion-like particle $A_y$ has been built-in as the extra-space component of the higher dimensional gauge field. Though $A_y$ is 
not a NG boson associated with a global symmetry, it has a shift symmetry associated with higher dimensional local gauge symmetry. 
The coupling of the axion-like particle $A_y$ with gluons is attributed to the ``Chern-Simons" (CS) term, which is radiatively induced. We adopted a toy model to calculate the CS term with some unknown gauge symmetry U(1)$_{\rm X}$. The CS term was obtained in two ways: first by a concrete 1-loop calculation and next by use of the Fujikawa's method \cite{Fujikawa} to deal with the chiral anomaly in 4D space-time. The obtained results are identical. We argued that this means that the radiative correction to the CS term is ``1-loop exact", since the result obtained by the latter method should be exact, valid in all orders of the perturbation. The Wilson coefficient of the resultant CS term is also free from UV-divergence even though the theory itself is non-renormalizable higher dimensional theory.

Though the scenario discussed in this paper has some similarity to those with axion fields, it also has its own novel feature. The loop-induced effective lagrangian, corresponding to the Chern-Simons term, obtained by summing up all KK modes in the internal state of the Feynman diagram, is no longer linear in the field $A_y$ as in the ordinary CS term but a periodic function of $A_y$. Actually this is an inevitable consequence of the higher dimensional gauge theory where the compactified extra space is non-simply-connected manifold such as $S^{1}$, since in such theories $A_y$ has a physical meaning as the phase of Wilson-loop. We argued how such novel feature of this scenario causes the modification of the solutions of the strong CP problem based on the axion fields and discussed how the mass scales of the theory, $M_c = 1/R$, $M$ are constrained by the astrophysical and cosmological observations.

Here is a remark. In the previous section we discussed only the P-conserving quadratic terms with respect to the pseudo-scalar fields $\hat{A}_y$ and $\pi^{0}$ in $V_{eff}$, eq.(\ref{6.10}). Actually, because of the fact that $f(A_y)$ is not linear in $A_y$ and the presence of the VEV $\bar{A}_y$, there also exist terms which are odd powers of the pseudo-scalar fields, such as $f^{(2)}(\bar{A}_y)\hat{A}_y^{2}\pi^{0}$, which violate P symmetry. Let us note that  $f^{(2n)}(0) = 0$ ($n$: positive integer) as $f(A_y)$ is an odd function of $A_y$, but $f^{(2n)}(\bar{A}_y)$ are generally non-vanishing. 
Namely, such P-violating effects are due to the VEV of the P-odd field $A_y$. Such higher order terms, however, are always accompanied by the higher inverse powers of $M_{c} = 1/R$, since $f(A_y)$ is the function of $2\pi gRA_y$, the phase of the Wilson-loop. Thus, these higher order terms are strongly suppressed by the inverse powers of $M_c$, which is expected to be of the order of, say, 10 TeV at least, and will be harmless. In addition, when the bulk mass $M$ of $\psi$ is larger than $M_{c}$, such higher order terms get exponential suppression by the powers of $e^{-2\pi RM} = e^{-2\pi M/M_{c}}$, because of the suppression of higher order derivatives $f^{(2n)}(\bar{A}_y)$, as we have seen in the text for the case of the de-compactification limit. 

For the purpose to clarify the essence of the mechanism to solve the strong CP problem in the framework of higher dimensional gauge theories, in this paper we adopted a toy model. In order to make the model realistic, there remain some issues to be settled. 
The SU(3)$_{\rm c} \times$ U(1)$_{X}$ model we discussed is a non-chiral (vector-like) theory, taking $S^{1}$ as the extra dimension. 
This choice, however, has a drawback. Namely, the KK zero-mode of the gluon's extra-space component $G^{a}_{y}$ remains as a massless state in the theory, which clearly contradicts with reality. A possible way out of this problem is to adopt an orbifold $S^{1}$/$Z_{2}$ as the extra dimension, instead of $S^{1}$, and assign the same overall $Z_{2}$-parity for all elements of the color triplets. Then $G^{a}_{y}$ has an odd $Z_2$-parity and does not have a massless KK zero-mode, while the KK zero-mode of the ordinary gluon $G^{a}_{\mu}$, having an even $Z_2$-parity, remains massless. On the other hand, because of the choice of the orbifold, fermions $u, d$ and $\psi$ all become Weyl fermions and the QCD becomes a chiral theory. We are thus enforced to double the matter fields as $d^{\alpha}_{1}, \ d^{\alpha}_{2}$, for instance, and assign opposite overall $Z_{2}$-parities for each of them, so that their KK zero-modes behave as $d^{\alpha}_{L}, \ d^{\alpha}_{R}$. 

In the case of the U(1)$_X$ gauge boson, the situation is just opposite. We need $A_y$, not $A_{\mu}$ as the member of the effective low-energy theory. A possible way to realize this situation is to embed U(1)$_X$ in some non-Abelian gauge group. The simplest candidate for the non-Abelian gauge group will be SU(2) \cite{Kubo} and we may regard the pair of $\psi^{\alpha}_{1}$, $\psi^{\alpha}_{2}$ to form a 
SU(2) doublet. Then $A_{M}$ is identified with a part of the SU(2) gauge boson, which connects the different elements of the doublet. Since, $\psi^{\alpha}_{1}$, $\psi^{\alpha}_{2}$ have opposite overall $Z_{2}$-parities, now $A_{\mu}$ has odd $Z_{2}$-parity and therefore its KK zero-mode disappears, while $A_y$, having even $Z_{2}$-parity, possesses a massless KK zero-mode, as is expected. 

By adopting the orbifold, ordinary bulk mass term is forbidden and should be replaced by so-called $Z_{2}$-odd bulk mass term, such as 
$- \epsilon (y) M (\overline{\psi^{\alpha}_{1}}\psi^{\alpha}_{1} + \overline{\psi^{\alpha}_{2}}\psi^{\alpha}_{2})$ ($\epsilon (y)$ takes either of $\pm$ 1 depending on the sign of $y$). Such mass term is known to lead to an exponential suppression of the KK zero-mode's mass. One may wonder whether this implies the possibility of a too light exotic colored fermion $\psi$, which clearly is undesirable. Fortunately, the 5D field $\psi$ obeys the anti-periodic boundary condition along the extra dimension and therefore does not possess the KK zero-mode. In any case, more complete analysis of such extended model is necessary, though we leave it for further investigation.

\subsection*{Acknowledgments}

We would like to thank K. Hasegawa for his contribution to this work until some stage of its progress. Thanks are also due to T. Misumi for useful and informative discussions. This work was supported in part by Japan Society for the Promotion of Science, Grants-in-Aid for Scientific Research, No.~15K05062.


\providecommand{\href}[2]{#2}\begingroup\raggedright\endgroup

\end{document}